# NaCl-assisted CVD growth of wafer scale high quality trilayer MoS$_2$ and the role of concentration boundary layer

*Aditya Singh[1], Madan Sharma[1] and Rajendra Singh[1,2]*

[1]Department of Physics, Indian Institute of Technology Delhi, New Delhi, India-110016

[2]Nanoscale Research Facility, Indian Institute of Technology Delhi, New Delhi, India-110016

## ABSTRACT

Direct growth of wafer scale high quality 2D layered materials (2DLMs) on SiO$_2$/Si substrate is still a challenge. The chemical vapor deposition (CVD) technique has played a significant role in achieving a large area continuous film of 2DLMs. CVD growth requires the optimization of many growth parameters such as temperature, amount of precursors, pressure, carrier gas flow and distance between the reactants. However, the role of boundary layer of reactants concentration has not been explored yet. The amount of precursors which leads to the formation of reactants concentration boundary layer has a significant role in controlling the thickness of growing material. Here, we report the role of concentration boundary layer to achieve wafer-scale MoS$_2$ in NaCl-assisted CVD growth at low temperature. Control of boundary layer thickness has led to the synthesis monolayer, bilayer, trilayer, and bulk MoS$_2$ film and flakes in our single-zone CVD at atmospheric pressure. Most importantly, we have synthesized 7 × 2.5 cm$^2$ area continuous, high quality trilayer MoS$_2$ film with good repeatability. We believe that our approach may lead to synthesize other wafer-scale 2DLMs that will pave the way for nano- and optoelectronics.

**KEYWORDS**

Large area growth, molybdenum disulfide, chemical vapor deposition, salt-assisted, concentration boundary layer, Raman spectroscopy, XPS

## 1. INTRODUCTION

Two-dimensional (2D) transition metal dichalcogenides (TMDs) materials are getting significant attention owing to their remarkable properties in monolayer (1L) and multilayer (ML)s forms[1]. Particularly, molybdenum disulfide ($MoS_2$), a family member of TMDs, has thickness dependent, bandgap (1.8 eV for 1L and 1.2 eV for bulk)[2], high mobility (40-480 $cm^2$/V-s)[3,4], and high Seebeck coefficient[5]. These unique properties make $MoS_2$ a fitting material candidate for flexible devices[4,6], photodetectors[7,8], field-effect transistors[4,9], space applications[10,11], gas sensors[12], etc. Various techniques have been utilized to synthesize layered $MoS_2$, such as metal-organic chemical vapor deposition (MOCVD)[13,14], chemical vapor deposition (CVD)[15–18], atomic layer deposition (ALD)[19], physical vapor deposition (PVD)[20], and thermal deposition[21]. Among them, CVD has been used considerably to synthesize large-scale, high-quality 1L-$MoS_2$ at high-temperatures (700-850°C)[22,23].

Achieving wafer-scale growth at low-temperature without compromising film quality is essential for industrial scale applications. So, the efficacy of various growth promoters/catalysis was investigated to minimize growth temperature and maximize area coverage. The addition of synergistic additives such as fluorides[24], alkali metal halide[13,25,26], KOH[27], and seed catalysis to the growth substrates leads to exponential growth area coverage[28,29]. Singh *et al.*[30] did low-temperature NaCl-assisted growth of 1L-$MoS_2$ on crystalline, amorphous, and layered substrates and observed that NaCl forms the layer of $Na_2S$ and/or $Na_2SO_4$ underneath of $MoS_2$ layer, which makes the layer transfer process very smooth[31]. Salt-assisted $MoS_2$ has shown high crystallinity, high mobility (~100 cm$^2$/V-s), and high optical properties matching with the level of conventional CVD $MoS_2$ monolayers[13].

In optimizing CVD growth for a material, essential parameters are growth temperature, precursors, pressure, carrier gas flow, the distance between precursors, growth substrate, etc. But the kinetics and mass-transport mechanisms of CVD growth are less explored. Mainly, the impact of the boundary layer on CVD growth has not been explored yet. As the amount of reactants is crucial in CVD growth, so the concentration boundary layer formation. So the understanding of the role of the concentration boundary layer is critical in achieving desired growth. In this paper, we have performed NaCl-assisted CVD growth of $MoS_2$ and played with the boundary layer to synthesize wafer-scale trilayer (3L) $MoS_2$ film. Utilizing the various characterization techniques, physical and chemical properties of the as-grown wafer-scale 3L-$MoS_2$ have been analyzed.

## 2. EXPERIMENTAL

In the present work, to achieve wafer-scale $MoS_2$ film, we have utilized the same CVD growth parameters for NaCl-assisted growth of $MoS_2$ as in our previous work[30]. However, to investigate the role of the concentration boundary layer on the CVD growth, we have tuned the distance between $MoO_3$+NaCl precursors and growing substrate. We have used $SiO_2$ (300 nm, thermally oxidized silicon)/Si as a growth substrate. Sulfur (S), sodium chloride (NaCl), and molybdenum trioxide ($MoO_3$) precursors procured from Sigma-Aldrich were used in a single-zone mini CVD setup (MTI Corporation, USA). Cleaning of the substrates was achieved by acetone followed by IPA and DI water. The amount of precursors was taken as, S = 100 mg, $MoO_3$ = 15 mg, and NaCl = 50 mg. The growth of $MoS_2$ film was achieved by mixing NaCl powder with $MoO_3$ powder in a crucible placed at the middle (hottest zone) of the CVD tube. In another crucible, sulfur powder was placed at a lower temperature zone at 14 cm away from the $MoO_3$+NaCl mixture. Preoccupied precursors, dust, moisture, and contamination were removed by passing argon (Ar) gas at 480 sccm for 10 minutes at 300 °C. After that Ar at 120 sccm was passed for the remaining CVD growth process. S started melting and vaporizing around 540±10 °C and 640± 5 °C, respectively. NaCl-assisted $MoS_2$ growth was obtained at 650°C for 10 min, and after that system was left for normal cooling.

For initial characterization, optical microscopy (OM) has been beautifully utilized as a quantitative technique for determining layer numbers as in graphene on $SiO_2$/Si[32]. So, we have used Nikon Eclipse LV100 OM to estimate layer thickness. Estimation of layer thickness was corroborated by Raman and photoluminescence spectroscopy using LabRAM HR Evolution (Horiba Scientific) with 514 nm laser at room temperature (RT). The surface topography of as-grown $MoS_2$ film was mapped by atomic force microscopy (AFM) (Agilent 5600 LS). Field emission scanning electron microscopy (FESEM) was done using Oxford-EDX system IE 250 X Max 80. X-ray photoelectron spectroscopy (XPS) was done by monochromatic Al Kα X-ray line (energy 1486.7 eV, probe size ~ 1.75 mm × 2.75 mm). The X-ray diffraction (XRD) measurements were performed with Cu Kα (λ= 1.54 Å) source using the Philips Xpert Pro system. Transmission electron microscope (TEM) images and selected area electron diffraction (SAED) were captured by FEI Tencai F20.

## 3.   RESULTS AND DISCUSSION

NaCl plays a considerable role in the CVD of TMDs by reducing the growth temperature[33,34] and streamlining the deposition of metal-oxyhalide species onto the growth substrate. Figure 1(a) shows the schematic illustration of a typical powder based CVD process for NaCl-assisted 3L-MoS$_2$ growth in a single zone CVD setup. Gaseous precursors form a concentration boundary layer under the streamlined flow of Ar (see Figure 1(c)). Under the laminar flow, the boundary layer is the region where the velocity of the reactants changes from zero at the wall to the maximum at the central axis of the tube. The gaseous precursors flowing above the boundary layer must diffuse through it and reach the deposition surface in order to start nucleation and growth. The formation of the concentration boundary layer is schematically shown in Figure 1(c), where from left to right, the concentration of precursors and their velocity are increasing. Inside the tube, yellow and white region represent laminar flow of the reactants and boundary layer region, respectively. The thickness of the boundary layer, $t$, is directly proportional to square root of the distance and inversely proportional to the square root of Reynolds number as following[35,36]:

$$t = \sqrt{\frac{x}{R_e}} \qquad (1)$$

where, $R_e = \rho u_x/\mu$, $\rho$= mass density, u= reactant flow velocity, $x =$ distance from the inlet in the flow direction, and $\mu$= viscosity.

The central axis of the CVD tube may be considered as the core axis of the boundary layer. Our CVD tube has an inner diameter of 4.4 cm so the central axis of concentration boundary can be considered at 2.2 cm from the inner surface of the tube. To understand the role of the boundary layer, let us define a few parameters such as $s$, $h$, and $d$ as the separation between the precursors and growing face of SiO$_2$/Si substrate, the height of the boat, and distance between the growing face of the Si substrate and central axis of the tube/ boundary layer, respectively. (see Figure 1(b)). Figure 1(d) depicts the different regions on the growing substrate. It has been observed that when the substrate was placed over the crucible, a region that comes in between the sidewalls of the crucible (point A) has a greater probability of having 3L-MoS$_2$ film than the region that is outside of the wall (point C). Point B show the interface region between MoS$_2$ film and SiO$_2$/Si substrate.

On changing the parameters *s, d* and *h*, the impact of the concentration boundary layer on MoS$_2$ film changes. We have chosen six cases/samples, namely, S1, S2, S3, S4, S5, and S6, of size ~ 1 × 2.5 cm$^2$. Figures 2 and 3, parameters *s* and *d* have been varied, and OM images at points A, B, and C for the corresponding sample have been captured. When the sample (S1) was near to MoO$_3$+NaCl mixture (*s*= 0.4 cm), and away from the concentration boundary layer (*d*=1.6 cm), OM images show mid of sample has bulk MoS$_2$ (point A, Figure 2(a)), the interface has multi-layer flakes (point B, Figure 2(b)), and point C show triangular flakes and film of 1L-MoS$_2$ (Figure 2(c)). It is obvious from Figure 1(b) that increasing *s* will decrease *d*. So, when *s* was increased to 0.7 cm, and *d* was reduced to 1.4 cm (sample S2), point A has bulk MoS$_2$ while point B has a few-layer MoS$_2$ film and point C has effectively no growth (see Figures 2(d)-(f)). Furthermore, at *s*=0.7 cm and *d*=1.2 cm, point A has multi-layer MoS$_2$ film and over which bulk crystals are observed (later on, these bulk crystals were found to be unreacted Mo (see supporting information (SI), Figure S6, and Table ST2). Initially, when the samples S1 and S2 were closer to MoO$_3$+NaCl mixture (high concentration), and away from the boundary layer, mid-region of samples is bulk MoS$_2$ while at relatively near location (d=1.2 cm), multi-layer MoS$_2$ film with Mo crystals were observed.

Furthermore, when sample S4 moved closer to the boundary layer (*d*=1.0 cm), the mid-region shows nucleation centers, the interface shows continuous film and flakes, and point C shows no growth (see Figures 3(a)-(c)). In case/sample S5, when *d*=0.4 cm, the mid-region shows continuous film (point A), interface shows continuous film too (point B), and point C shows no growth (see Figures 3(d)-(f)). The quality of as-grown MoS$_2$ continuous film is very high (confirmed by several measurements and shown in Figures 5-7) and the number of layers found to be three. It seems that the sample is close enough to the boundary layer so that passing sulfur atoms can reach the deposition surface quickly and far enough from the mixture so that only the required amount of $MoO_3+NaCl$ can reach the growing surface, and they will not form bulk $MoS_2$. When the sample is further moved closer to the boundary layer (sample S6, $d$=0.2 cm), due to high flux of S precursors and Ar gas, white round patches and contamination over the mid-area and obviously no growth at the interface and point C is observed. It appears that S6 was confined within the boundary layer and a high flux of Ar atoms sputtered the sample after the growth. Comparing samples S1 to S6, S5 is found to be the highest quality continuous 3L-$MoS_2$.

Growth by the same parameters as used in S5 has been further extended on $7 \times 2.5$ cm$^2$ substrate (a whole circular Si wafer could not be inserted due to the small diameter of our CVD tube, so a rectangular wafer was inserted). Remarkably, full area coverage of continuous film of 3L-$MoS_2$ was obtained (see Figure 4(a)), and the zoomed-in view is shown in Figure 4(b). This sample was characterized by Raman, AFM, X-ray diffraction (XRD), TEM, etc. (see SI Figures S1-S5). The quality of the as-grown film is found to be very high. However, some white patches were observed on the surface (see a white region in Figure 4(a)), and FE-SEM and EDAX analysis show that they are unreacted cuboidal crystals of Mo (see SI S5). As the growing face of the substrate is placed in the face-down manner, it is expected those excess Mo atoms from $MoO_3$ also be adsorbed over the substrate rather than diffusing as byproducts.

To resolve this issue, we placed five ~ 3.0 cm$^2$ substrates altogether over the crucible so that there is some gap (at micro-level) between two substrates. As expected, continuous film of 3L-$MoS_2$ on the whole wafer, without any white patches, is grown (see Figure 4(c)). Figures 4(e), (f) are zoomed-in view of OM images of Figure 4(c), and they show no contamination/impurity and grain boundary formation (at the micro-level) on the as-grown film. Figure 4(d) depicts the FE-SEM image of 3L-MoS2 taken from the marked region in Figure 4(b). Clearly, it has no grain boundary formation even at the nano-level. Figures 4(g) and (h) show schematics of the effect of separation between $MoO_3$ + NaCl precursors and growing substrate and impact on growth by placing a single large-sized wafer versus many substrates altogether, respectively. In the following passages, the as-grown wafer-scale 3L-$MoS_2$ film will be characterized to analyze its physical and chemical properties.

To investigate the morphology and measure the surface potential of the as-grown 3L-$MoS_2$ film, AFM and Kelvin probe force microscopy (KPFM) were performed, respectively (see Figure 5). The calculated root-mean-square (RMS) roughness of 3L-$MoS_2$ over 100 $\mu m^2$ area is 0.91 nm. Using the KPFM technique, the contact potential difference ($V_{CPD}$) distribution 3L-$MoS_2$ was observed to be Gaussian in nature with FWHM of about 30 meV. $V_{CPD}$ between the tip and the sample is defined as

$$e V_{CPD} = \phi_{tip} - \phi_{sample} \qquad (2)$$

where, $\phi_{tip}$ and $\phi_{sample}$ are the work function of tip and sample, respectively. $\phi_{tip}$ was calculated by considering highly oriented pyrolytic graphite (HOPG) as a reference sample whose work function was taken to be 4.6 eV[37,38]. So the measured $V_{CPD}$ between the tip (Pt/Ir) and the HOPG was 0.297 V that leads to $\phi_{tip}$ to be 4.897 eV. Furthermore, the measured $V_{CPD}$ of 3L-$MoS_2$ film was -4.2 mV and calculated $\phi_{MoS_2}$ was found to be 4.9 eV.

Raman scattering measurements on $MoS_2$ film were carried out by 514 nm laser line at the room-temperature to observe phonon vibrations. $MoS_2$ molecule has two first-order fundamental Raman modes, $E_{2g}^1/E$ (in-plane vibration) and $A_{1g}/A$ (out-of-plane vibration)[39]. The position of $E_{2g}^1$ and $A_{1g}$ modes depend upon layer number, strain, and impurities, etc[40,41]. Figure 6(a) shows the typical Raman spectrum of 3L-MoS$_2$ measured at the region marked in Figure 4(c) where $E_{2g}^1 \sim 384.5$ cm$^{-1}$ and $A_{1g} \sim 408.1$ cm$^{-1}$, and peak position difference ($\Delta\omega = \omega_A - \omega_E$) is $\sim 23.6$ cm$^{-1}$, which clearly indicates formation of 3L-MoS$_2$[42]. Figure 6(b) shows spatial distribution of $E_{2g}^1$ and $A_{1g}$ modes, measured over the various positions on the sample as shown in inset schematic. As deviation in peak position of $E_{2g}^1$ and $A_{1g}$ modes is $\leq 0.5$ cm$^{-1}$ (resolution of the spectrometer), so it indicates as-grown 3L-MoS$_2$ has high uniformity over wafer-scale. Furthermore, integrated intensity ratio of $A_{1g}$ and $E_{2g}^1$ Raman modes ($\frac{I_A}{I_E}$) and their peak position separation $\Delta\omega$ ($= \omega_A - \omega_E$) can also be considered as parameters to verify uniformity. In Figure 6 (c), $\Delta\omega$ and ($\frac{I_A}{I_E}$) are calculated from the data of Figure 6 (b) and these values are nearly constant over the substrate, hence showing the uniformity of the as-grown 3L-MoS$_2$ over wafer-scale.

As 3L-MoS$_2$ has indirect bandgap, so photoluminescence cannot be done to measure the bandgap, hence, UV-visible spectroscopy in reflectance mode was carried out to achieve this (see inset of Figure 6(d)). The bandgap (E$_g$) was calculated by translating reflectance UV-vis curve to Kubelka-Munk (K-M) function F(R) and plotting $[F(R)h\nu]^2$ versus photon energy (h$\nu$) by using the well-known equation,

$$[F(R)h\nu]^2 = B(h\nu - E_g) \qquad (3)$$

where B is a constant. The bandgap of 3L-MoS$_2$ was calculated to be $1.72 \pm 0.01$ eV, matching with the previously reported value[43].

XPS measurements were carried out to investigate the chemical properties of as-grown 3L-MoS$_2$. Figure 7(a) shows the XPS survey spectrum of 3L-MoS$_2$ and Figures 7(b) and 7(c) show the core level XPS spectrum of molybdenum and sulfur, respectively. The binding energy (BE) of C 1$s$ peak was taken at 284.8 eV for the calibration purpose. The BE of core level orbitals of Mo$^{4+}$, i.e., 3d$_{5/2}$ and 3d$_{3/2,}$ are observed at 229.9 and 233.0 eV, respectively, with FWHM of ~1.7 eV, which are in agreement with the earlier reports[44]. S 2$p$ XPS spectrum depicts spin-orbit doublets, 2$p_{3/2}$ and 2$p_{1/2}$, positioned at 163.0 and 164.3 eV, respectively, indicating the energy separation of 1.3 eV.

X-ray diffraction (XRD) measurements on 3L-MoS$_2$ were performed in and the recorded diffraction pattern is shown in Figure 7(d). Peaks centered at 13.9°, 33.0° , and 69.0° correspond to MoS$_2$, Si (002), and Si(400), respectively. It can be clearly seen that as-grown 3L-MoS$_2$ shows single crystallinity. To further confirm it, transmission electron microscopy (TEM) was done by transferring 3L-MoS$_2$ on the TEM grid (see Figure 7(e)). Selected area electron diffraction (SAED) pattern was recorded at the marked rectangle in TEM image and shown in Figure 7(f). SAED pattern depicts single-crystalline nature of grown MoS$_2$ film whose interplanar spacing d is 0.29 nm, consisting with the previous reports[22].

## 4. CONCLUSIONS

In conclusion, we have synthesized high-quality wafer-scale 3L-MoS$_2$ by controlling the concentration boundary layer. The physical and chemical properties of 3L-MoS$_2$ have been analyzed by optical, Raman, AFM, XPS, XRD, TEM, etc. Controlling the concentration boundary layer formation by tuning separation between the precursors and growing face of the substrate; layer number, growth area coverage, and nucleation density can be regulated easily, as shown in optical images. High uniformity and low roughness of wafer-scale 3L-$MoS_2$ were confirmed by Raman scattering and AFM measurements, respectively. UV-vis spectroscopy in reflectance mode was carried to calculate the bandgap of wafer-scale $MoS_2$ film, and it was found to be 1.72 eV, indicating the trilayer nature of $MoS_2$. Our as-grown 3L-$MoS_2$ films are single crystalline as identified by XRD and TEM measurements. We believe that regulated control over concentration boundary can lead to high-quality wafer-scale growth of other TMDs for engineering future nanoelectronics.

**ASSOCIATED CONTENT**

**Supporting Information**

Raman, AFM, XRD, TEM, and EDAX data of 3L-$MoS_2$ on single wafer is provided.

**AUTHOR INFORMATION**

**Corresponding Author**

**Rajendra Singh-** [1]Department of Physics, Indian Institute of Technology Delhi, New Delhi, India-110016

[2]Nanoscale Research Facility, Indian Institute of Technology Delhi, New Delhi, India-110016
Email: rsingh@physics.iitd.ac.in;

**Author**

**Aditya Singh-** [1]Department of Physics, Indian Institute of Technology Delhi, New Delhi, India-110016;

**Funding-** This work is partially supported by the Grand Challenge Project "MBE growth of 2D-Materials" under grant No. MI10800

**Notes-** The authors declare no competing financial interest.

**ACKNOWLEDGMENT**

The authors are grateful to the Central Research Facility (CRF) at the Indian Institute of Technology Delhi for providing the characterization facilities. A. S. would like to thank University

Grant Commission (UGC) for providing the research fellowship. This work is partially supported by the Grand Challenge Project "MBE growth of 2D-Materials" under grant No. MI10800, funded by MHRD and IIT Delhi.

# REFERENCES

(1) Wang, Q. H.; Kalantar-Zadeh, K.; Kis, A.; Coleman, J. N.; Strano, M. S. Electronics and Optoelectronics of Two-Dimensional Transition Metal Dichalcogenides. *Nature Nanotechnology* **2012**, *7* (11), 699–712. https://doi.org/10.1038/nnano.2012.193.

(2) Kuc, A.; Zibouche, N.; Heine, T. Influence of Quantum Confinement on the Electronic Structure of the Transition Metal Sulfide TS2. *Phys. Rev. B* **2011**, *83* (24), 245213. https://doi.org/10.1103/PhysRevB.83.245213.

(3) Bao, W.; Cai, X.; Kim, D.; Sridhara, K.; Fuhrer, M. S. High Mobility Ambipolar MoS2 Field-Effect Transistors: Substrate and Dielectric Effects. *Appl. Phys. Lett.* **2013**, *102* (4), 042104. https://doi.org/10.1063/1.4789365.

(4) Shao, P.-Z.; Zhao, H.-M.; Cao, H.-W.; Wang, X.-F.; Pang, Y.; Li, Y.-X.; Deng, N.-Q.; Zhang, J.; Zhang, G.-Y.; Yang, Y.; Zhang, S.; Ren, T.-L. Enhancement of Carrier Mobility in MoS2 Field Effect Transistors by a SiO2 Protective Layer. *Appl. Phys. Lett.* **2016**, *108* (20), 203105. https://doi.org/10.1063/1.4950850.

(5) Buscema, M.; Barkelid, M.; Zwiller, V.; van der Zant, H. S. J.; Steele, G. A.; Castellanos-Gomez, A. Large and Tunable Photothermoelectric Effect in Single-Layer MoS2. *Nano Lett.* **2013**, *13* (2), 358–363. https://doi.org/10.1021/nl303321g.

(6) Liu, K.; Yan, Q.; Chen, M.; Fan, W.; Sun, Y.; Suh, J.; Fu, D.; Lee, S.; Zhou, J.; Tongay, S.; Ji, J.; Neaton, J. B.; Wu, J. Elastic Properties of Chemical-Vapor-Deposited Monolayer MoS2, WS2, and Their Bilayer Heterostructures. *Nano Lett.* **2014**, *14* (9), 5097–5103. https://doi.org/10.1021/nl501793a.

(7) Moun, M.; Singh, A.; Tak, B. R.; Singh, R. Study of Photoresponse Behavior of High Barrier Pd/MoS2/Pd Photodetector. *J. Phys. D: Appl. Phys.* **2019**. https://doi.org/10.1088/1361-6463/ab1f59.

(8) Lopez-Sanchez, O.; Lembke, D.; Kayci, M.; Radenovic, A.; Kis, A. Ultrasensitive Photodetectors Based on Monolayer MoS2. *Nature Nanotechnology* **2013**, *8* (7), 497–501. https://doi.org/10.1038/nnano.2013.100.

(9) Peelaers, H.; Van de Walle, C. G. Effects of Strain on Band Structure and Effective Masses in MoS2. *Phys. Rev. B* **2012**, *86* (24), 241401. https://doi.org/10.1103/PhysRevB.86.241401.

(10) Vogl, T.; Sripathy, K.; Sharma, A.; Reddy, P.; Sullivan, J.; Machacek, J. R.; Zhang, L.; Karouta, F.; Buchler, B. C.; Doherty, M. W.; Lu, Y.; Lam, P. K. Radiation Tolerance of Two-Dimensional Material-Based Devices for Space Applications. *Nat Commun* **2019**, *10* (1), 1202. https://doi.org/10.1038/s41467-019-09219-5.

(11) Singh, A.; Singh, R. γ-Ray Irradiation-Induced Chemical and Structural Changes in CVD Monolayer MoS2. *ECS J. Solid State Sci. Technol.* **2020**, *9* (9), 093011. https://doi.org/10.1149/2162-8777/abb583.

(12) Shokri, A.; Salami, N. Gas Sensor Based on MoS2 Monolayer. *Sensors and Actuators B: Chemical* **2016**, *236*, 378–385. https://doi.org/10.1016/j.snb.2016.06.033.

(13) Kim, H.; Ovchinnikov, D.; Deiana, D.; Unuchek, D.; Kis, A. Suppressing Nucleation in Metal–Organic Chemical Vapor Deposition of MoS$_2$ Monolayers by Alkali Metal Halides. *Nano Letters* **2017**, *17* (8), 5056–5063. https://doi.org/10.1021/acs.nanolett.7b02311.

(14) Li, H.; Li, Y.; Aljarb, A.; Shi, Y.; Li, L.-J. Epitaxial Growth of Two-Dimensional Layered Transition-Metal Dichalcogenides: Growth Mechanism, Controllability, and Scalability. *Chem. Rev.* **2018**, *118* (13), 6134–6150. https://doi.org/10.1021/acs.chemrev.7b00212.

(15) Zhou, J.; Lin, J.; Huang, X.; Zhou, Y.; Chen, Y.; Xia, J.; Wang, H.; Xie, Y.; Yu, H.; Lei, J.; Wu, D.; Liu, F.; Fu, Q.; Zeng, Q.; Hsu, C.-H.; Yang, C.; Lu, L.; Yu, T.; Shen, Z.; Lin, H.; Yakobson, B. I.; Liu, Q.; Suenaga, K.; Liu, G.; Liu, Z. A Library of Atomically Thin Metal Chalcogenides. *Nature* **2018**, *556* (7701), 355. https://doi.org/10.1038/s41586-018-0008-3.

(16) Shi, Y.; Li, H.; Li, L.-J. Recent Advances in Controlled Synthesis of Two-Dimensional Transition Metal Dichalcogenides via Vapour Deposition Techniques. *Chemical Society Reviews* **2015**, *44* (9), 2744–2756. https://doi.org/10.1039/C4CS00256C.

(17) Singh, A.; Moun, M.; Singh, R. Effect of Different Precursors on CVD Growth of Molybdenum Disulfide. *Journal of Alloys and Compounds* **2019**, *782*, 772–779. https://doi.org/10.1016/j.jallcom.2018.12.230.

(18) Lee, Y.-H.; Zhang, X.-Q.; Zhang, W.; Chang, M.-T.; Lin, C.-T.; Chang, K.-D.; Yu, Y.-C.; Wang, J. T.-W.; Chang, C.-S.; Li, L.-J.; Lin, T.-W. Synthesis of Large-Area MoS2 Atomic Layers with Chemical Vapor Deposition. *Advanced Materials* **2012**, *24* (17), 2320–2325. https://doi.org/10.1002/adma.201104798.

(19) Jurca, T.; Moody, M. J.; Henning, A.; Emery, J. D.; Wang, B.; Tan, J. M.; Lohr, T. L.; Lauhon, L. J.; Marks, T. J. Low-Temperature Atomic Layer Deposition of MoS2 Films.

*Angewandte Chemie International Edition* **2017**, *56* (18), 4991–4995. https://doi.org/10.1002/anie.201611838.

(20) Tumino, F.; Casari, C. S.; Passoni, M.; Russo, V.; Bassi, A. L. Pulsed Laser Deposition of Single-Layer MoS2 on Au(111): From Nanosized Crystals to Large-Area Films. *Nanoscale Adv.* **2019**, *1* (2), 643–655. https://doi.org/10.1039/C8NA00126J.

(21) Liu, K.-K.; Zhang, W.; Lee, Y.-H.; Lin, Y.-C.; Chang, M.-T.; Su, C.-Y.; Chang, C.-S.; Li, H.; Shi, Y.; Zhang, H.; Lai, C.-S.; Li, L.-J. Growth of Large-Area and Highly Crystalline MoS2 Thin Layers on Insulating Substrates. *Nano Lett.* **2012**, *12* (3), 1538–1544. https://doi.org/10.1021/nl2043612.

(22) Tao, L.; Chen, K.; Chen, Z.; Chen, W.; Gui, X.; Chen, H.; Li, X.; Xu, J.-B. Centimeter-Scale CVD Growth of Highly Crystalline Single-Layer MoS2 Film with Spatial Homogeneity and the Visualization of Grain Boundaries. *ACS Appl. Mater. Interfaces* **2017**, *9* (13), 12073–12081. https://doi.org/10.1021/acsami.7b00420.

(23) Yang, P.; Zou, X.; Zhang, Z.; Hong, M.; Shi, J.; Chen, S.; Shu, J.; Zhao, L.; Jiang, S.; Zhou, X.; Huan, Y.; Xie, C.; Gao, P.; Chen, Q.; Zhang, Q.; Liu, Z.; Zhang, Y. Batch Production of 6-Inch Uniform Monolayer Molybdenum Disulfide Catalyzed by Sodium in Glass. *Nature Communications* **2018**, *9* (1), 979. https://doi.org/10.1038/s41467-018-03388-5.

(24) Liu, C.; Xu, X.; Qiu, L.; Wu, M.; Qiao, R.; Wang, L.; Wang, J.; Niu, J.; Liang, J.; Zhou, X.; Zhang, Z.; Peng, M.; Gao, P.; Wang, W.; Bai, X.; Ma, D.; Jiang, Y.; Wu, X.; Yu, D.; Wang, E.; Xiong, J.; Ding, F.; Liu, K. Kinetic Modulation of Graphene Growth by Fluorine through Spatially Confined Decomposition of Metal Fluorides. *Nature Chemistry* **2019**, *11* (8), 730–736. https://doi.org/10.1038/s41557-019-0290-1.

(25) Wang, Z.; Xie, Y.; Wang, H.; Wu, R.; Nan, T.; Zhan, Y.; Sun, J.; Jiang, T.; Zhao, Y.; Lei, Y.; Yang, M.; Wang, W.; Zhu, Q.; Ma, X.; Hao, Y. NaCl-Assisted One-Step Growth of MoS$_2$ –WS $_2$ in-Plane Heterostructures. *Nanotechnology* **2017**, *28* (32), 325602. https://doi.org/10.1088/1361-6528/aa6f01.

(26) Barreau, N.; Bernède, J. C.; Pouzet, J.; Guilloux-Viry, M.; Perrin, A. Characteristics of Photoconductive MoS2 Films Grown on NaCl Substrates by a Sequential Process. *physica status solidi (a)* **2001**, *187* (2), 427–437. https://doi.org/10.1002/1521-396X(200110)187:2<427::AID-PSSA427>3.0.CO;2-I.

(27) Zhu, J.; Xu, H.; Zou, G.; Zhang, W.; Chai, R.; Choi, J.; Wu, J.; Liu, H.; Shen, G.; Fan, H. MoS2–OH Bilayer-Mediated Growth of Inch-Sized Monolayer MoS2 on Arbitrary Substrates. *J. Am. Chem. Soc.* **2019**, *141* (13), 5392–5401. https://doi.org/10.1021/jacs.9b00047.

(28) Yang, P.; Yang, A.-G.; Chen, L.; Chen, J.; Zhang, Y.; Wang, H.; Hu, L.; Zhang, R.-J.; Liu, R.; Qu, X.-P.; Qiu, Z.-J.; Cong, C. Influence of Seeding Promoters on the Properties of CVD Grown Monolayer Molybdenum Disulfide. *Nano Res.* **2019**, *12* (4), 823–827. https://doi.org/10.1007/s12274-019-2294-y.

(29) Li, S.; Lin, Y.-C.; Zhao, W.; Wu, J.; Wang, Z.; Hu, Z.; Shen, Y.; Tang, D.-M.; Wang, J.; Zhang, Q.; Zhu, H.; Chu, L.; Zhao, W.; Liu, C.; Sun, Z.; Taniguchi, T.; Osada, M.; Chen, W.; Xu, Q.-H.; Wee, A. T. S.; Suenaga, K.; Ding, F.; Eda, G. Vapour–Liquid–Solid Growth of Monolayer MoS 2 Nanoribbons. *Nature Materials* **2018**, *17* (6), 535–542. https://doi.org/10.1038/s41563-018-0055-z.

(30) Singh, A.; Moun, M.; Sharma, M.; Barman, A.; Kumar Kapoor, A.; Singh, R. NaCl-Assisted Substrate Dependent 2D Planar Nucleated Growth of MoS2. *Applied Surface Science* **2021**, *538*, 148201. https://doi.org/10.1016/j.apsusc.2020.148201.

(31) Sharma, M.; Singh, A.; Singh, R. Monolayer MoS2 Transferred on Arbitrary Substrates for Potential Use in Flexible Electronics. *ACS Appl. Nano Mater.* **2020**, *3* (5), 4445–4453. https://doi.org/10.1021/acsanm.0c00551.

(32) Blake, P.; Hill, E. W.; Castro Neto, A. H.; Novoselov, K. S.; Jiang, D.; Yang, R.; Booth, T. J.; Geim, A. K. Making Graphene Visible. *Applied Physics Letters* **2007**, *91* (6), 063124. https://doi.org/10.1063/1.2768624.

(33) Li, S.; Wang, S.; Tang, D.-M.; Zhao, W.; Xu, H.; Chu, L.; Bando, Y.; Golberg, D.; Eda, G. Halide-Assisted Atmospheric Pressure Growth of Large WSe2 and WS2 Monolayer Crystals. *Applied Materials Today* **2015**, *1* (1), 60–66. https://doi.org/10.1016/j.apmt.2015.09.001.

(34) Lee, D. K.; Kim, S.; Oh, S.; Choi, J.-Y.; Lee, J.-L.; Yu, H. K. Water-Soluble Epitaxial NaCl Thin Film for Fabrication of Flexible Devices. *Scientific Reports* **2017**, *7* (1). https://doi.org/10.1038/s41598-017-09603-5.

(35) Handbook of Chemical Vapor Deposition (CVD) | ScienceDirect https://www.sciencedirect.com/book/9780815514329/handbook-of-chemical-vapor-deposition-cvd#book-info (accessed Feb 22, 2021).

(36) Ohring, M. *Engineering Materials Science*; Academic Press, 1995.

(37) Melitz, W.; Shen, J.; Lee, S.; Lee, J. S.; Kummel, A. C.; Droopad, R.; Yu, E. T. Scanning Tunneling Spectroscopy and Kelvin Probe Force Microscopy Investigation of Fermi Energy Level Pinning Mechanism on InAs and InGaAs Clean Surfaces. *Journal of Applied Physics* **2010**, *108* (2), 023711. https://doi.org/10.1063/1.3462440.

(38) Beerbom, M. M.; Lägel, B.; Cascio, A. J.; Doran, B. V.; Schlaf, R. Direct Comparison of Photoemission Spectroscopy and in Situ Kelvin Probe Work Function Measurements on Indium Tin Oxide Films. *Journal of Electron Spectroscopy and Related Phenomena* **2006**, *152* (1), 12–17. https://doi.org/10.1016/j.elspec.2006.02.001.

(39) Li, H.; Zhang, Q.; Yap, C. C. R.; Tay, B. K.; Edwin, T. H. T.; Olivier, A.; Baillargeat, D. From Bulk to Monolayer MoS2: Evolution of Raman Scattering. *Advanced Functional Materials* **2012**, *22* (7), 1385–1390. https://doi.org/10.1002/adfm.201102111.

(40) Chae, W. H.; Cain, J. D.; Hanson, E. D.; Murthy, A. A.; Dravid, V. P. Substrate-Induced Strain and Charge Doping in CVD-Grown Monolayer MoS $_2$. *Applied Physics Letters* **2017**, *111* (14), 143106. https://doi.org/10.1063/1.4998284.

(41) Buscema, M.; Steele, G. A.; van der Zant, H. S. J.; Castellanos-Gomez, A. The Effect of the Substrate on the Raman and Photoluminescence Emission of Single-Layer MoS2. *Nano Research* **2014**, *7* (4), 561–571. https://doi.org/10.1007/s12274-014-0424-0.

(42) Yan, J.; Xia, J.; Wang, X.; Liu, L.; Kuo, J.-L.; Tay, B. K.; Chen, S.; Zhou, W.; Liu, Z.; Shen, Z. X. Stacking-Dependent Interlayer Coupling in Trilayer MoS2 with Broken Inversion Symmetry. *Nano Lett.* **2015**, *15* (12), 8155–8161. https://doi.org/10.1021/acs.nanolett.5b03597.

(43) Naik, M. H.; Jain, M. Origin of Layer Dependence in Band Structures of Two-Dimensional Materials. *Phys. Rev. B* **2017**, *95* (16), 165125. https://doi.org/10.1103/PhysRevB.95.165125.

(44) Kim, I. S.; Sangwan, V. K.; Jariwala, D.; Wood, J. D.; Park, S.; Chen, K.-S.; Shi, F.; Ruiz-Zepeda, F.; Ponce, A.; Jose-Yacaman, M.; Dravid, V. P.; Marks, T. J.; Hersam, M. C.; Lauhon, L. J. Influence of Stoichiometry on the Optical and Electrical Properties of Chemical Vapor Deposition Derived MoS2. *ACS Nano* **2014**, *8* (10), 10551–10558. https://doi.org/10.1021/nn503988x.

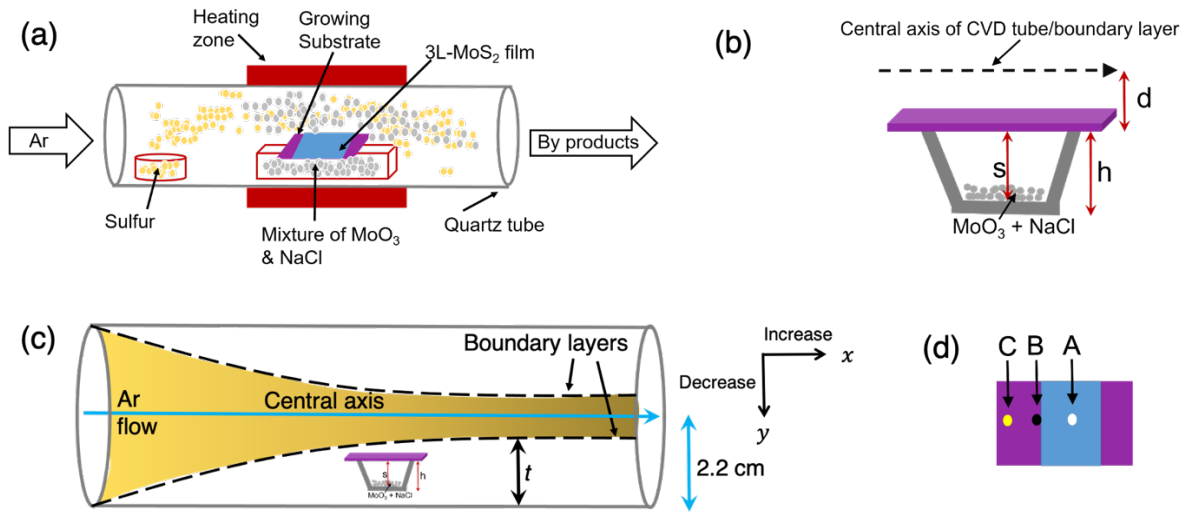

**Figure 1.** (a) Schematic illustration of typical CVD process for NaCl-assisted 3L-MoS$_2$ growth in single-zone CVD setup and (b) definition of *s*, *h*, and *d*. (c) Schematic of the reactant concentration gradient and its boundary layer formation in CVD tube. Yellow and white region inside the tube represent laminar flow of the reactants and boundary layer region, respectively. The concentration of sulfur is increasing in the direction of flow. (d) Schematic illustration of A, B, and C point on the substrate over which optical images are recorded in Figures 2 and 3.

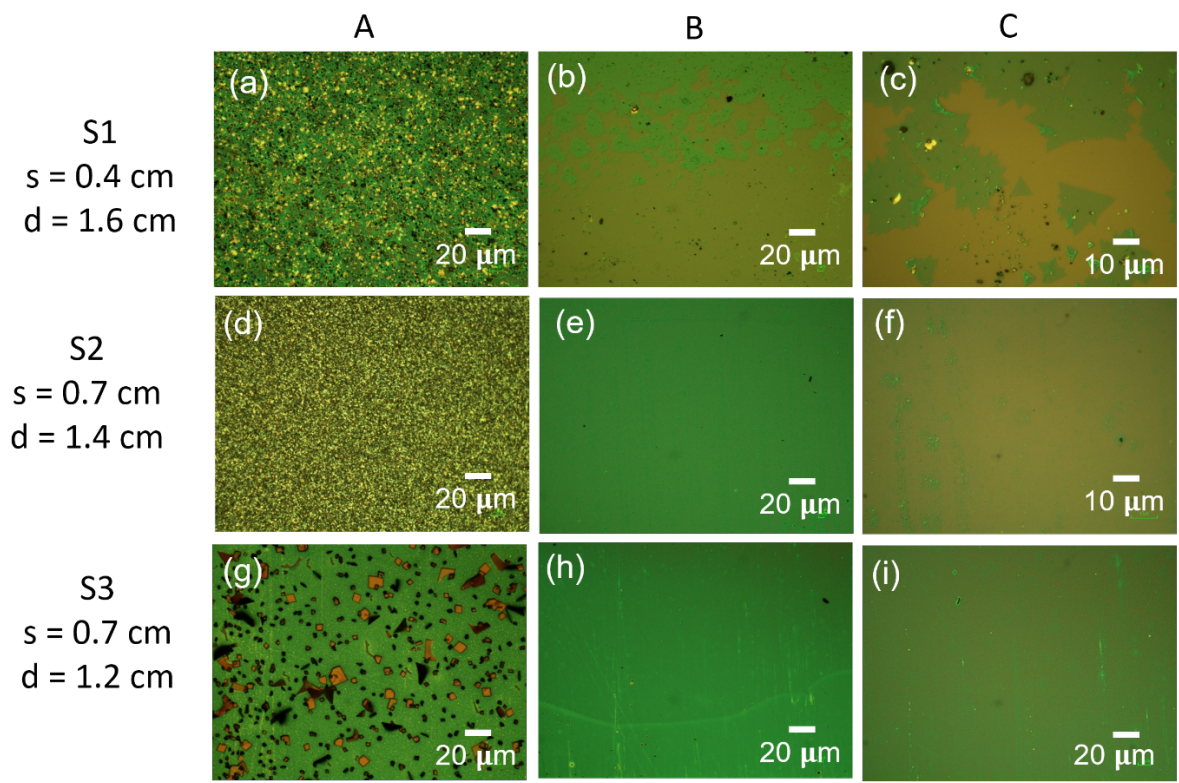

**Figure 2.** Optical images of the evolution of as-grown MoS₂ over the points A, B, and C on the substrates as the s and d are changing. Definition of points A, B, and C is shown schematically in figure 1 (d).

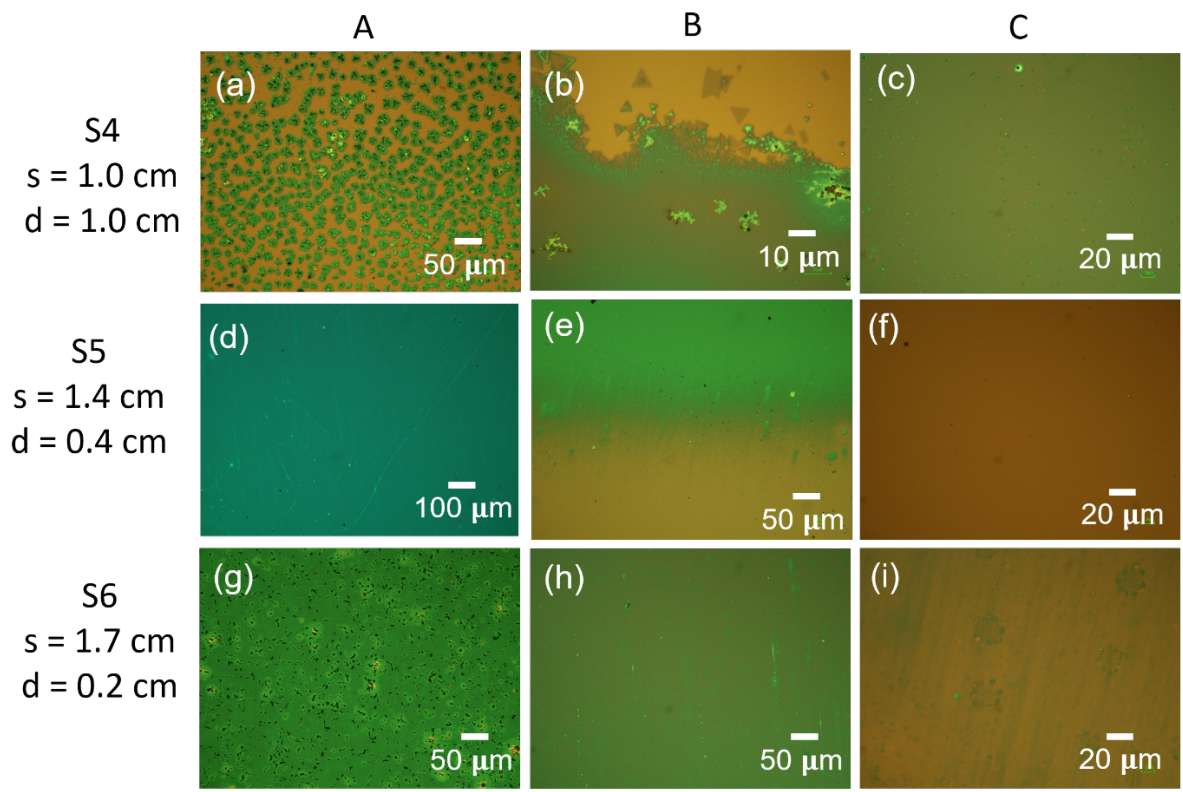

**Figure 3.** Optical images of the evolution of as-grown MoS$_2$ over the points A, B, and C on the substrates as the s and d are changing.

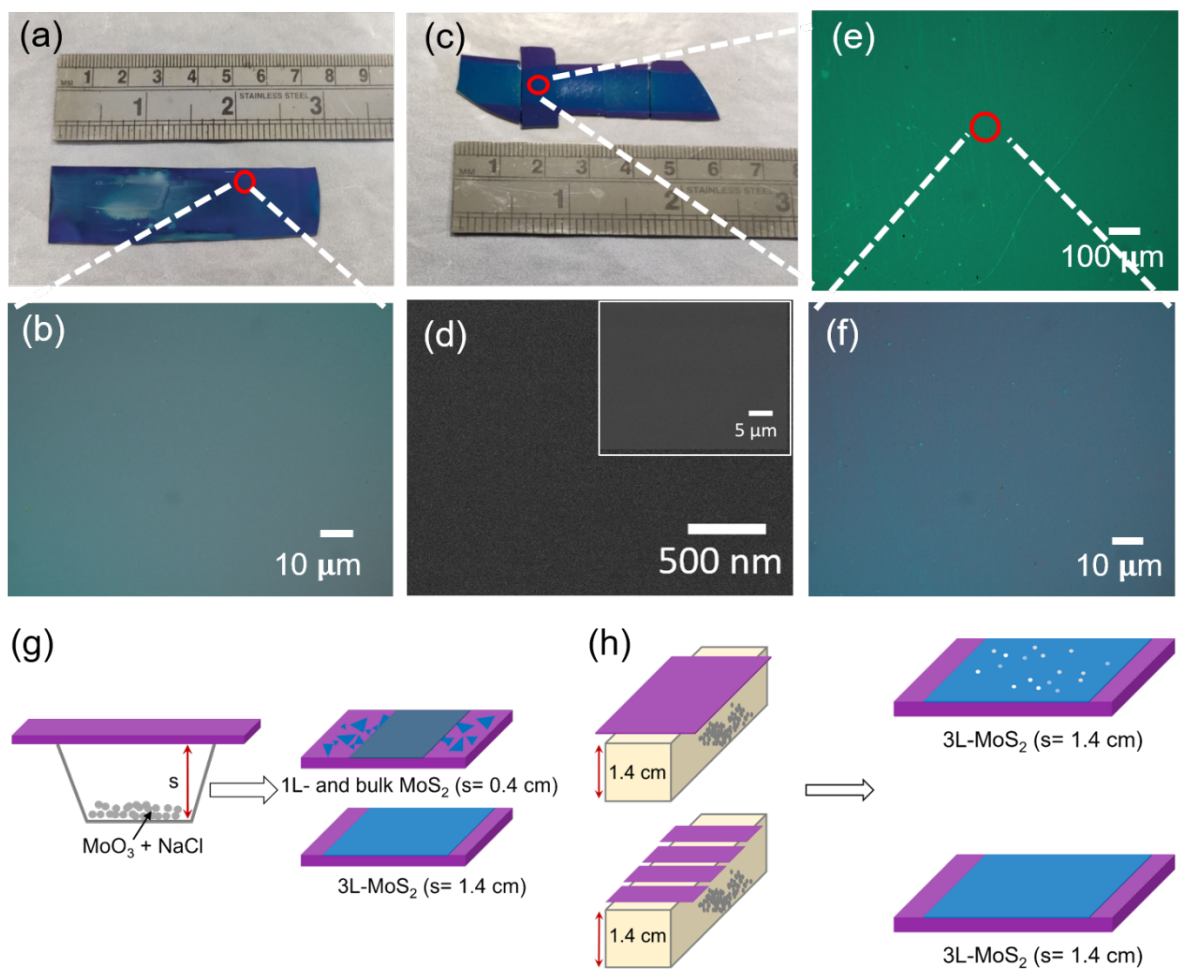

**Figure 4.** (a) Picture of as-grown 3L-MoS$_2$ when single SiO$_2$/Si substrate of $7 \times 2.5$ cm$^2$ area was placed. The white region shows unreacted molybdenum crystals, and corresponding EDAX measurements data is provided in supplementary information (b) Optical image taken over the marked region in (a). (c) Picture of as-grown 3L-MoS$_2$ when five SiO$_2$/Si substrates (area $\geq$ 2.5 cm$^2$) were placed alongside. (d) Field-emission scanning electron microscopy (FE-SEM) image of 3L-MoS$_2$ taken from the marked region in (b). 3L-MoS$_2$ shows no grain boundary at the micro- and nano-level. (e) Optical image of 3L-MoS$_2$ captured from the marked region in (b). (f) An enlarged optical image is taken from the marked area in (e). Schematics of (g) effect of separation between MoO$_3$ + NaCl precursors and growing substrate and (h) impact on growth by placing a single large-sized wafer versus many substrates altogether.

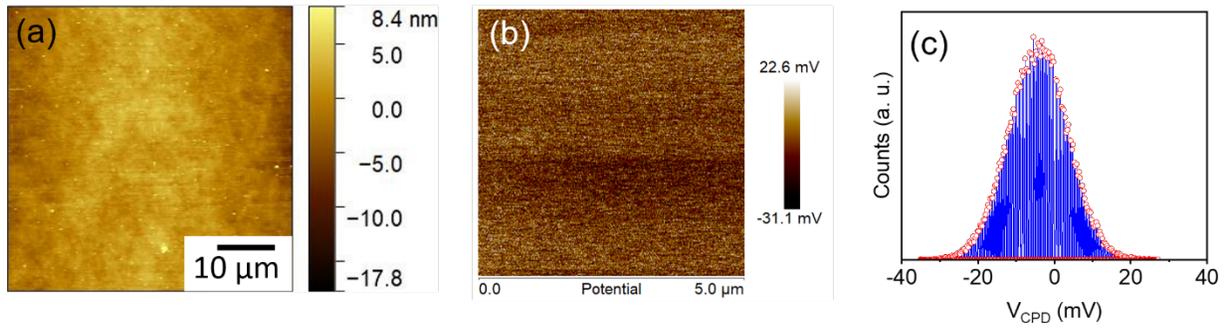

**Figure 5.** (a) Atomic force microscopy (AFM) image and (b) surface potential mapping captured by Kelvin probe force microscopy (KPFM) image of 3L-MoS$_2$ captured from the marked region in Figure 2(c). Gaussian distribution of contact potential difference (V$_{CPD}$) of 3L-MoS$_2$.

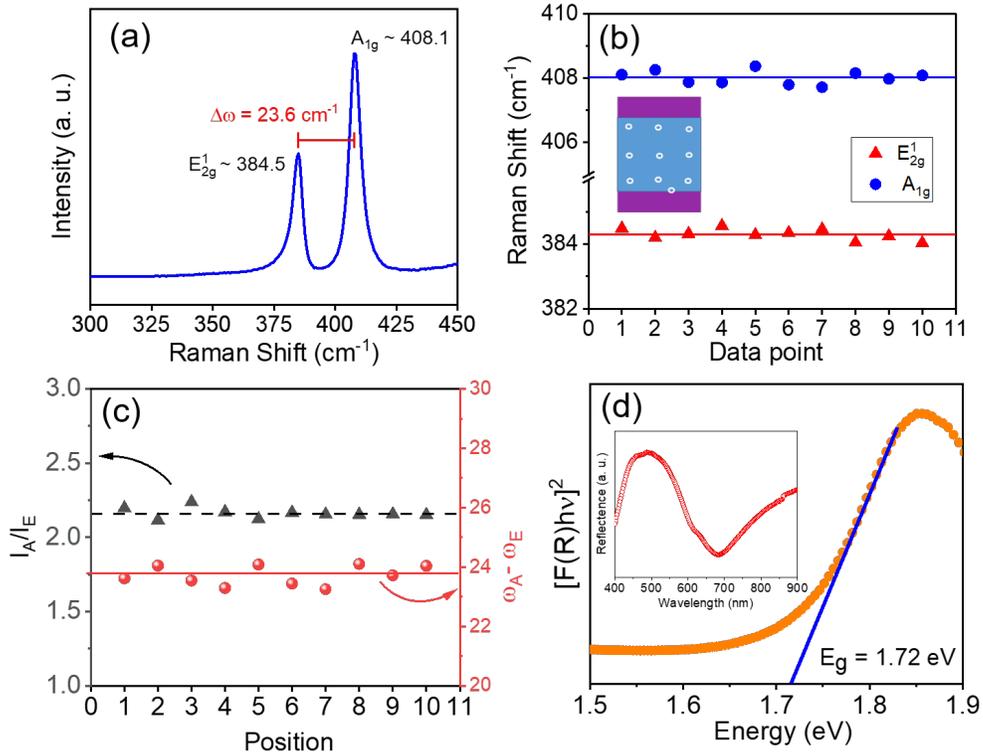

**Figure 6.** (a) Raman spectrum of 3L-MoS$_2$ measured at region marked in Figure 4(c). (b) Spatial distribution of $E_{2g}^1$ and $A_{1g}$ Raman modes of MoS$_2$, measured over the various positions on the sample as shown in the inset schematic. (c) The integrated intensity ratio of $A_{1g}$ and $E_{2g}^1$ Raman modes ($\frac{I_A}{I_E}$) and their peak position separation $\Delta\omega = \omega_A - \omega_E$. (d) Spectrum between $[F(R)h\nu]^2$ versus photon energy ($h\nu$) for 3L-MoS$_2$, where bandgap was calculated by translating reflectance curve (inset) to Kubelka-Munk (K-M) function F(R).

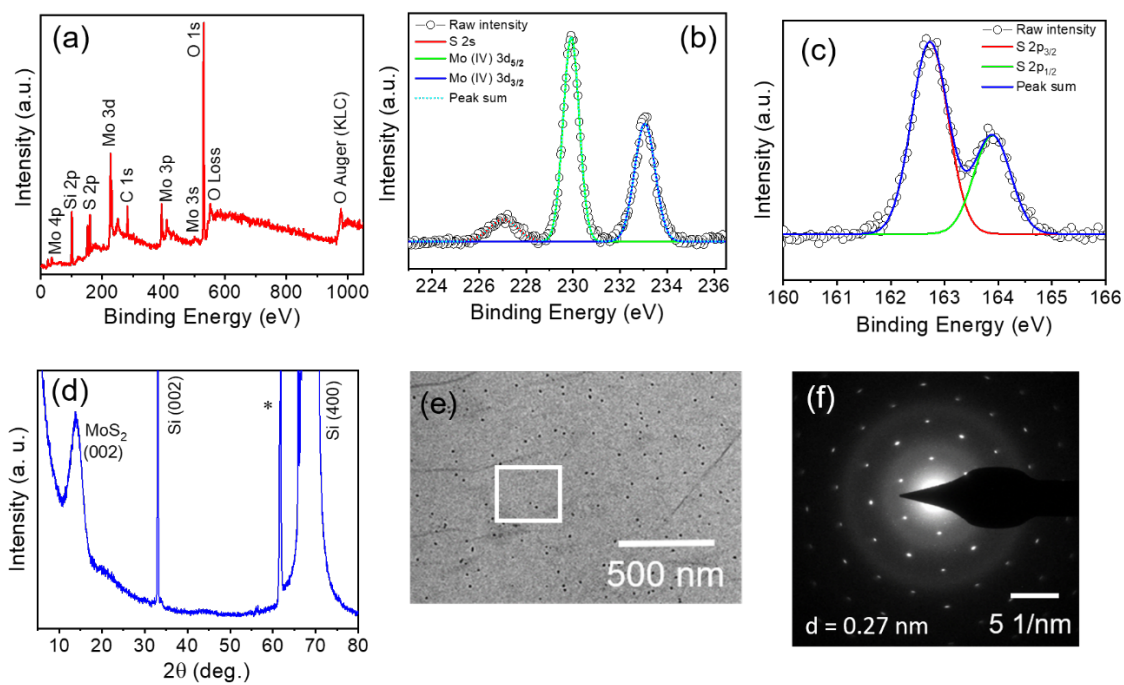

**Figure 7.** (a) X-ray photoelectron spectrum (XPS) survey spectrum of 3L-MoS$_2$. Core level XPS spectrum of (b) molybdenum and (c) sulfur. (d) X-ray diffraction (XRD) of 3L-MoS$_2$ grown over SiO$_2$/Si substrate. (e) TEM image of 3L-MoS$_2$ and marked white triangle show region where selected area electron diffraction (SAED) pattern was taken and shown in (f).

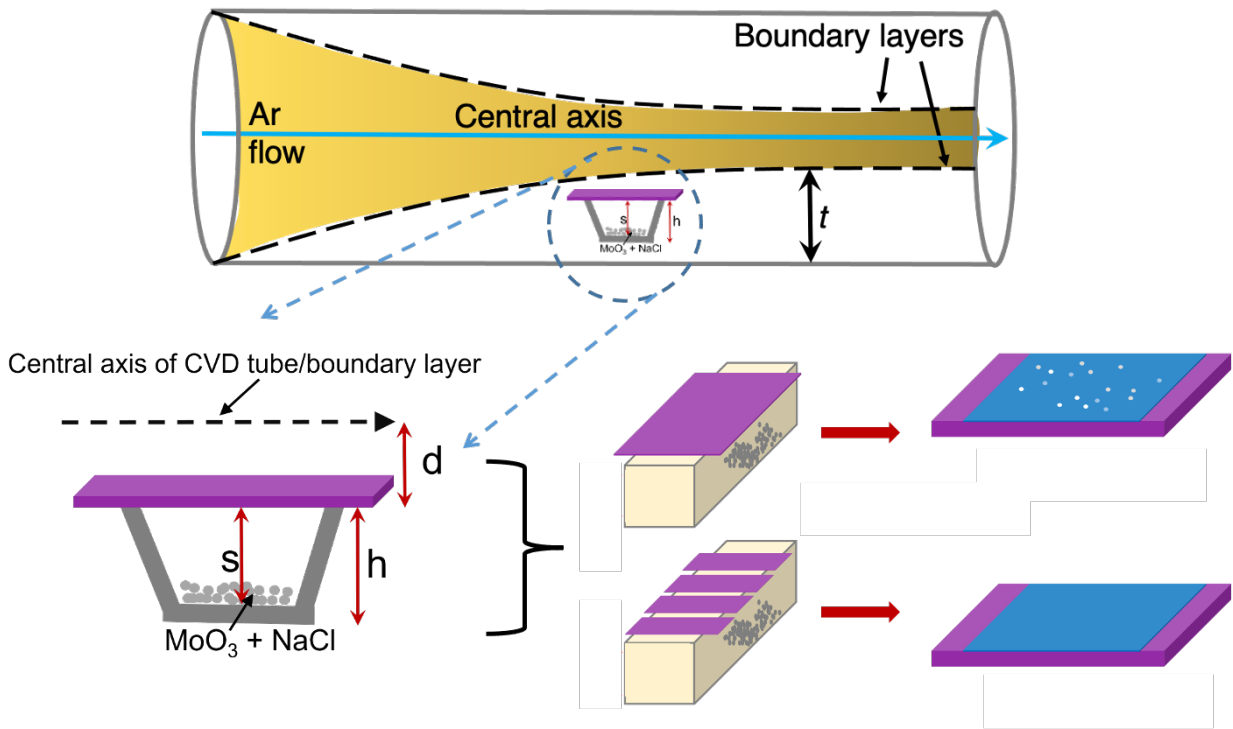

# Supplementary Information

# NaCl-assisted CVD growth of wafer scale high quality trilayer MoS$_2$ and the role of concentration boundary layer

*Aditya Singh[1], Madan Sharma[1] and Rajendra Singh[1,2]*

[1]Department of Physics, Indian Institute of Technology Delhi, New Delhi, India-110016
[2]Nanoscale Research Facility, Indian Institute of Technology Delhi, New Delhi, India-110016

S1: Raman scattering measurements

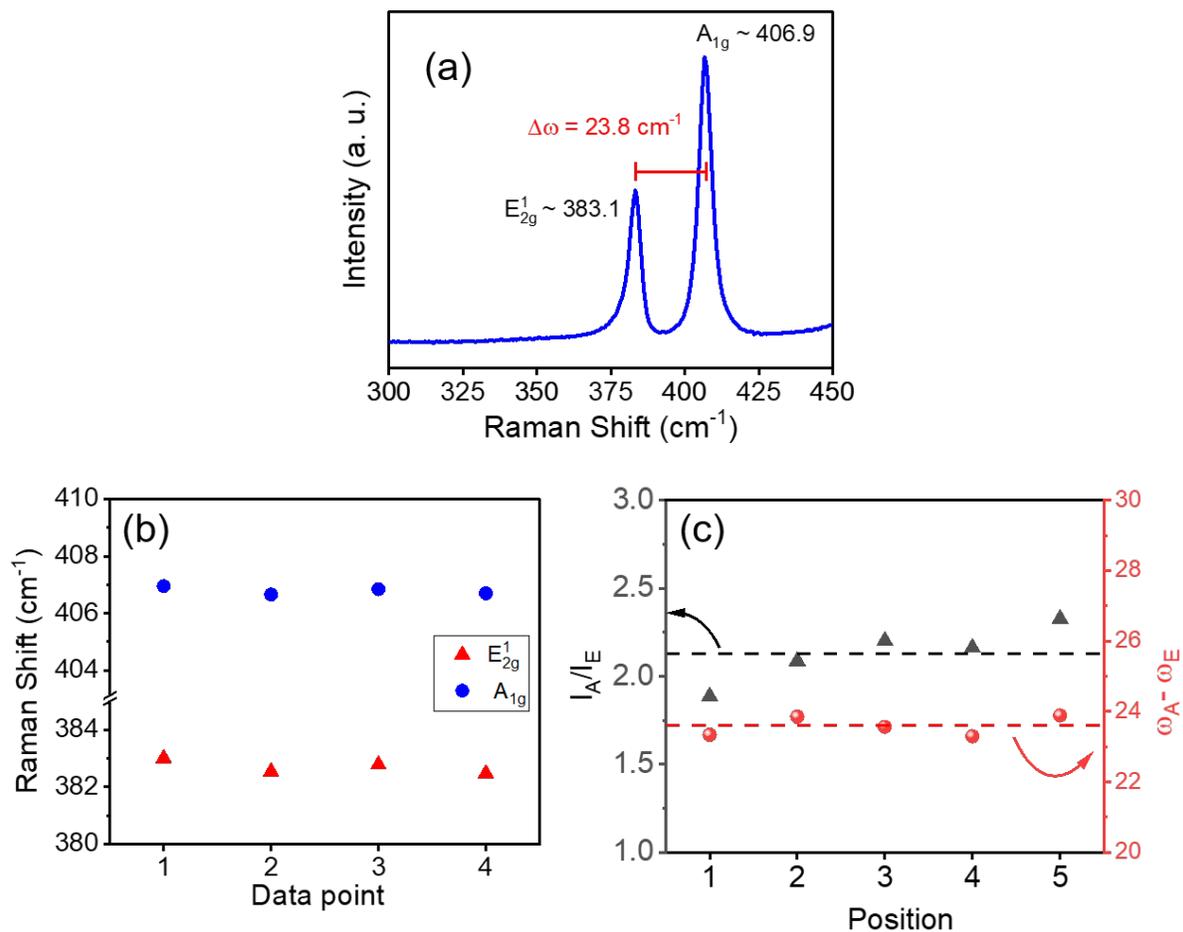

Figure S1: (a) Raman spectrum of 3L-MoS$_2$ measured at region marked in Figure 4(a). (b) Spatial distribution of $E^1_{2g}$ and $A_{1g}$ Raman modes of MoS$_2$, measured over the various positions on the sample. (c) Integrated intensity ratio of $A_{1g}$ and $E^1_{2g}$ Raman modes ($\frac{I_A}{I_E}$) and their separation $\Delta\omega = \omega_A - \omega_E$.

S2: Atomic force microscopy (AFM) measurements

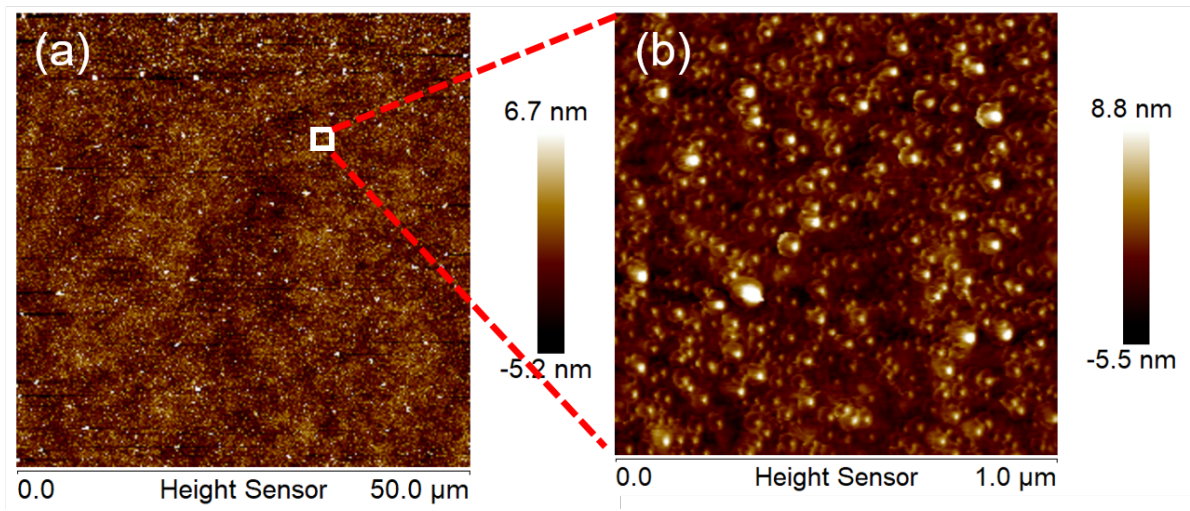

Figure S2: Atomic force microscopy (AFM) image of 3L-MoS$_2$ captured from marked region in Figure 4(a).

S3: X-ray diffraction (XRD) measurements

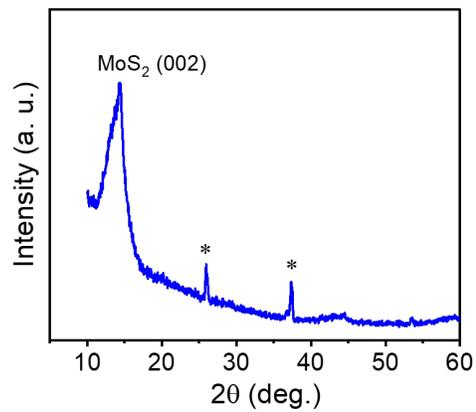

Figure S3: X-ray diffraction (XRD) of 3L-MoS$_2$ of sample shown in Figure 4(a).

S4: Transmission electron microscopy (TEM)

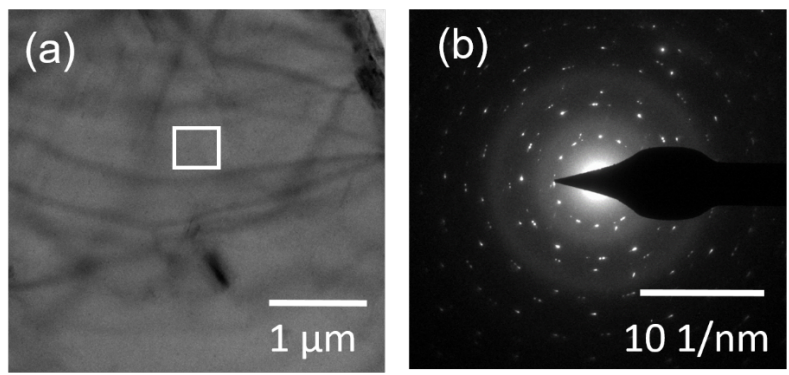

Figure S4: (a) TEM image of 3L-MoS$_2$ of sample shown in Figure 4(a) and marked white triangle show region where selected area electron diffraction (SAED) pattern was taken and shown in (b).

S5: Field emission scanning electron microscopy (FE-SEM) and energy dispersive X-ray analysis (EDAX)

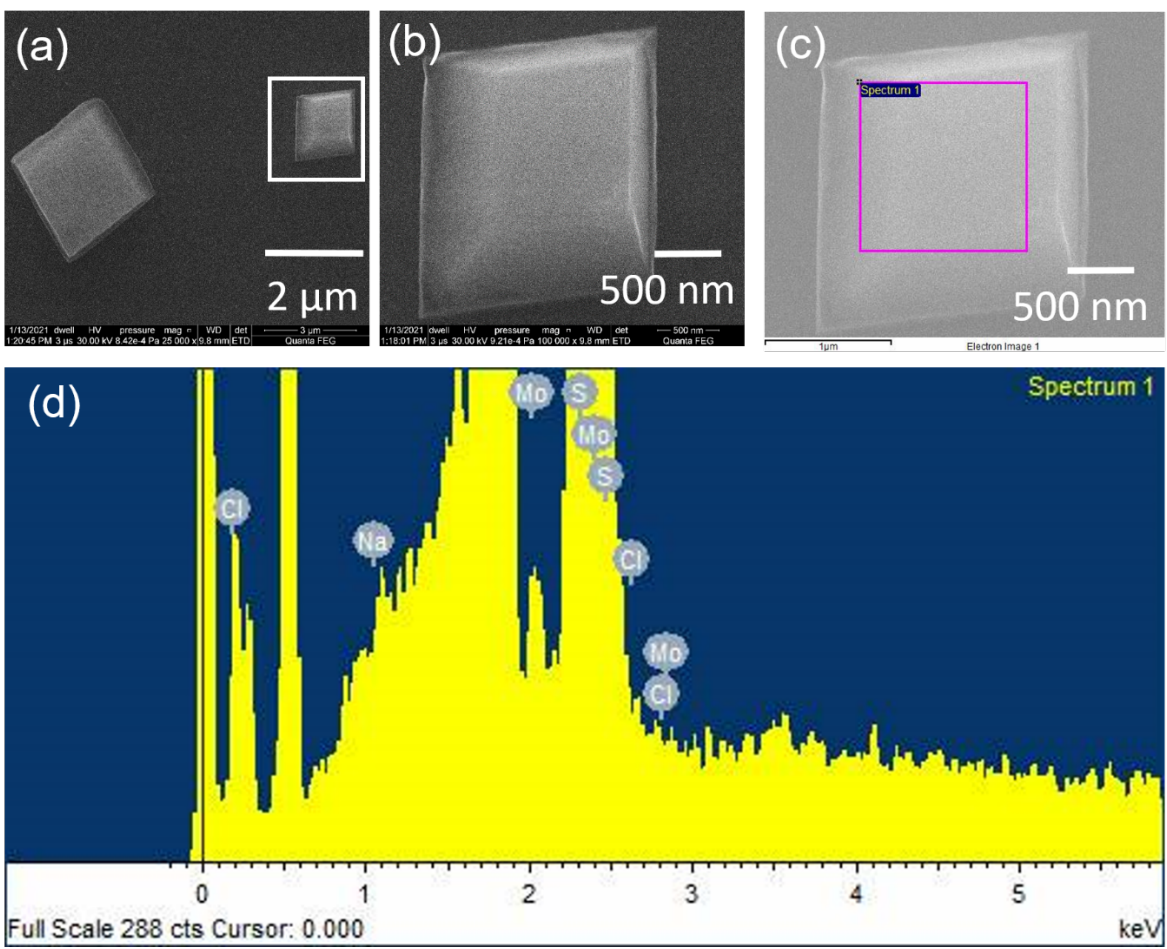

Figure S5: (a) FE-SEM image of salt look like molybdenum crystals left after the CVD growth. These crystals are found over the sample for which d= 0.4 cm (Figure 4(a)). (b) Enlarged view of crystal of (a). (c) Marked square show region where EDAX measurements are done.

Table ST1: Elemental composition of the crystal shown in Figure S5 (c).

| Element | Weight % | Atomic % |
| --- | --- | --- |
| Na | 0.00 | 0.00 |
| S | 0.00 | 0.00 |
| Cl | 0.00 | 0.00 |
| Mo | 100.00 | 100.00 |
| Total | 100.00 | |

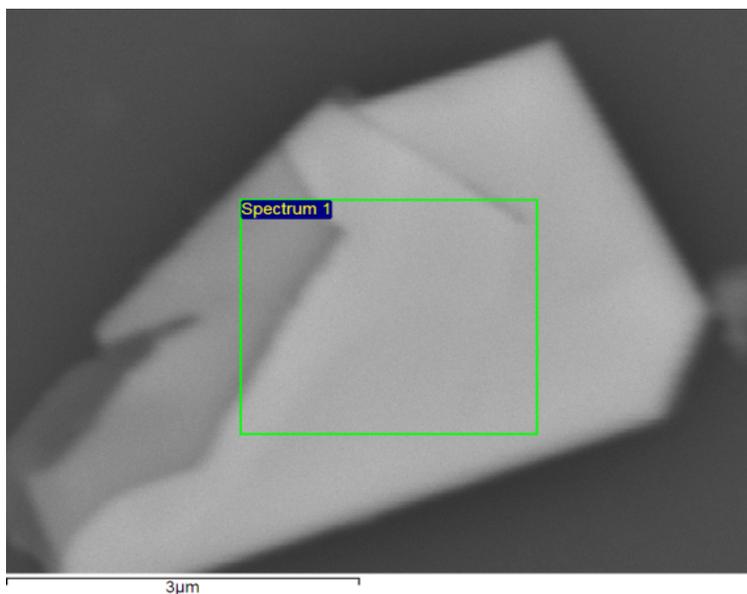

Figure S6: FE-SEM image of molybdenum crystals left after the CVD growth. These crystals are found over the sample for which d= 1.2 cm (Figure 2(g)).

Table ST2: Elemental composition of the crystal shown in Figure S6.

| Element | Weight % | Atomic % |
|---------|----------|----------|
| Na | 0.00 | 0.00 |
| S | 0.00 | 0.00 |
| Cl | 0.734 | 1.962 |
| Mo | 99.266 | 98.038 |
| Total | 100.00 | |

S7: Optical image of crucibles (boats) used in the growth

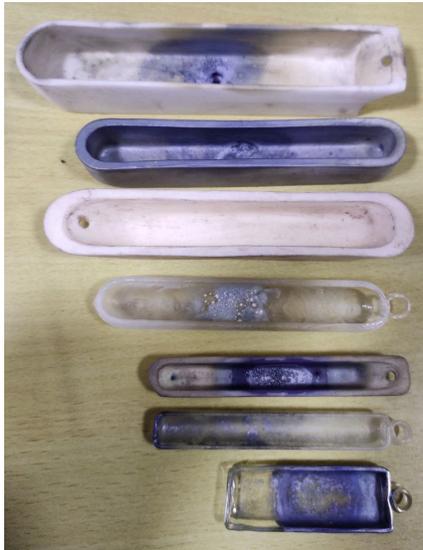

Figure S7: Optical image of crucibles (boats) used in the synthesis of MoS$_2$ for controlling the *s*, *d* and *h* to play with concentration boundary layer.